\documentclass[twocolumn]{aastex701}
\usepackage{graphicx}
\usepackage{amsmath}	
\usepackage{amssymb}	
\usepackage{cases}
\usepackage{array,multirow}
\usepackage{upgreek}
\usepackage{textcomp}
\usepackage{lineno}
\usepackage{subcaption}
\usepackage{mhchem}
\usepackage{longtable}
\usepackage{hyperref}
\usepackage[referable]{threeparttablex}
\usepackage{booktabs, dcolumn}

\newcommand{\Msol}{\,\mathrm{M}_\odot}

\newcommand{\au}{\,\texttt{au}}

\newcommand{\Teff}{T_{\mathrm{eff}}}
\newcommand{\Ledd}{L_{\mathrm{Edd}}}

\shorttitle{Pulsating Quasi-stars}
\shortauthors{Cantiello et al.}

\begin{document}

\title{Pulsational Instability of Quasi-Stars: \\ Interpreting the Variability of Little Red Dots}
\correspondingauthor{Matteo Cantiello}
\email{mcantiello@flatironinstitute.org}

\author[0000-0002-8171-8596]{Matteo Cantiello} 
\affil{Center for Computational Astrophysics, Flatiron Institute, 162 5th Avenue, New York, NY 10010, USA} 
\affil{Department of Astrophysical Sciences, Princeton University, Princeton, NJ 08544, USA}
\email{mcantiello@flatironinstitute.org}
\author[0009-0004-5685-6155]{Jake B. Hassan}
\affil{Department of Physics and Astronomy, Stony Brook University, Stony Brook, NY 11794-3800, USA}
\email{jakebhassan@gmail.com}
\author[0000-0002-3635-5677]{Rosalba Perna}
\affil{Department of Physics and Astronomy, Stony Brook University, Stony Brook, NY 11794-3800, USA}
\email{rosalba.perna@stonybrook.edu}
\author[0000-0001-5032-1396]{Philip J. Armitage}
\affil{Center for Computational Astrophysics, Flatiron Institute, 162 5th Avenue, New York, NY 10010, USA}
\affil{Department of Physics and Astronomy, Stony Brook University, Stony Brook, NY 11794-3800, USA}
\email{philip.armitage@stonybrook.edu}
\author[0000-0003-0936-8488]{Mitchell C. Begelman}
\affiliation{JILA, University of Colorado and National Institute of Standards and Technology, 440 UCB, Boulder, CO 80309-0440, USA}
\affiliation{Department of Astrophysical and Planetary Sciences, University of Colorado, 391 UCB, Boulder, CO 80309-0391, USA}
\email{mitch@jila.colorado.edu}
\author[0000-0002-2624-3399]{Yan-Fei Jiang}
\affil{Center for Computational Astrophysics, Flatiron Institute, 162 5th Avenue, New York, NY 10010, USA}
\email{jiang@flatitoninstitute.org}
\author[0000-0003-2012-5217]{Taeho Ryu}
\affiliation{JILA, University of Colorado and National Institute of Standards and Technology, 440 UCB, Boulder, CO 80309-0440, USA}
\affiliation{Department of Astrophysical and Planetary Sciences, University of Colorado, 391 UCB, Boulder, CO 80309-0391, USA}
\email{taeho.ryu@colorado.edu}
\affiliation{Max-Planck-Institut für Astrophysik, Karl-Schwarzschild-Straße 1, 85748 Garching bei München, Germany}
\author[0000-0002-2522-8605]{Richard H. D. Townsend}
\affiliation{ Department of Astronomy, University of Wisconsin-Madison, 475 N Charter St, Madison, WI, 53706, USA}
\email{townsend@astro.wisc.edu }

\begin{abstract}
The JWST discovery of ``Little Red Dots'' (LRDs) has revealed a population of compact, red sources at $z \sim 5$--$10$ that likely host supermassive black holes. Recent observations of the gravitationally lensed LRD R2211-RX1 reveal century-scale photometric variability and a hysteresis loop in the luminosity-temperature plane, strongly suggesting that the optical emission originates from a pulsating, stellar-like photosphere rather than an accretion disk. This supports the ``quasi-star'' hypothesis, where a rapidly growing black hole seed is embedded within a massive, radiation-pressure supported envelope. Here, we investigate the pulsational stability of these envelopes using the stellar evolution code \texttt{MESA} coupled with the non-adiabatic oscillation code \texttt{GYRE}. We identify a theoretical ``Quasi-Star Instability Strip'' with a blue edge at $T_{\mathrm{eff}} \approx 5000-5200$\,K. Models hotter than this threshold are stable, consistent with the non-variable LRD R2211-RX2 ($T_{\mathrm{eff}} \approx 5000$\,K), while cooler models are unstable to radial pulsations driven by the $\kappa$-mechanism in helium and hydrogen ionization zones. For quasi-star masses in the range $M_{\star}\sim10^4$--$10^5 \Msol$, we find that the unstable fundamental radial modes ($\ell =0$, n$_{\rm p}=1$) have periods  in the range $\sim 20$--$180$ years. The first overtone ($\ell =0$, n$_{\rm p}=2$) is also unstable or marginally stable in some of our models, with typical pulsation timescales $\sim 10$--$30$~years. These oscillations match the co-moving frame variability timescale of RX1. We argue that these violent pulsations likely drive enhanced mass loss analogous to super-AGB winds,  which could affect the duration of the quasi-star phase and regulate the final mass of the seeded black hole.
\end{abstract}


\section{Introduction} \label{sec:intro}

The discovery of high-redshift quasars powered by supermassive black holes with masses $M_{\rm BH} \gtrsim 10^9 \Msol$ less than 1 Gyr after the Big Bang \citep{Fan2006,Banados2018} remains a stringent test for models of cosmic structure formation. To reach such masses, stellar-mass seeds would require sustained super-Eddington accretion. Alternatively, ``heavy seeds'' ($10^4$--$10^6 \Msol$) formed via direct collapse could bypass the early bottleneck of exponential growth \citep{Begelman2006,Lodato2006}. 

A key prediction of the heavy seed scenario is the existence of ``quasi-stars'': massive, hydrostatic gas envelopes supported by the accretion luminosity of an embedded central black hole \citep{Begelman2008}. In this phase, the envelope acts as a calorimeter, reprocessing the hard accretion radiation into a cooler, thermal blackbody spectrum. While early models suggested the black hole could only grow to a small fraction of the envelope mass \citep{Ball2011,Ball2012}, recent refinements to the inner boundary conditions have shown that the black hole can essentially consume the entire envelope, growing to several tens of percent of the quasi-star mass
before the structure dissipates \citep{Coughlin2024, Hassan2025,Santarelly2025a}.

Observational support for this scenario has surged with the JWST discovery of ``Little Red Dots'' (LRDs)—compact, high-redshift objects characterized by a V-shaped spectral energy distribution (SED) \citep{Labbe2023,Matthee2024, Barro2024,Greene2024, Kokorev2024}. These objects exhibit broad Balmer lines indicative of high-velocity gas but lack the X-ray emission typical of unobscured AGN \citep{Maiolino2024}. \citet{Santarelli2025} and \citet{Begelman2025} demonstrated that the SEDs of LRDs are well-matched by late-stage quasi-stars on the Hayashi track, with effective temperatures $T_{\mathrm{eff}} \approx 4000$--$6000$\,K and radii $R \sim 1000\au$.

The spectra of LRDs are consistent with formation within a region of low gas density, but this is a prediction of both quasi-star and super-Eddington accretion models \citep{Liu2025}. Variability provides a route to determining whether the envelope is in (approximate) hydrostatic equilibrium, as expected uniquely for quasi-stars. \citet{Zhang2025} reported the discovery of a quadruply lensed LRD R2211-RX1 (RX1) at redshift $z\approx4.3$, consistent with exhibiting periodic variability over a 130-year baseline. The object traces a counter-clockwise hysteresis loop in the luminosity-temperature plane, a classic signature of stellar pulsation (the ``heat engine'' cycle) rather than the stochastic variability associated with accretion disks. The inferred period in the co-moving frame ($\sim 30$ years) and photospheric radius ($\sim 2000 \au$) are remarkably consistent with the dynamical timescales of the massive envelopes predicted by quasi-star theory. This result is particularly interesting, since LRDs do not appear to pulsate on timescales less than a year \citep{Inayoshi2025}. This apparent lack of short variability may disfavor the AGN scenario \citep{Secunda2025}. 

If quasi-stars are indeed pulsating, this has profound implications for their evolution. Previous studies have largely treated the envelope as static or subject only to continuum-driven winds \citep{Fiacconi2017, Hassan2025}. However, in evolved massive stars (e.g., AGB stars, Red Supergiants), pulsations can be a primary driver of mass loss \citep[see e.g.][]{YoonCantiello2010}. If similar physics applies to quasi-stars, pulsation-driven superwinds could drastically affect the duration of the LRD phase.

In this paper, we extend the structural models of \citet{Hassan2025} to the dynamical regime. We perform a linear stability analysis of quasi-star envelopes using the \texttt{GYRE} code to identify unstable modes and driving mechanisms. We then employ high-resolution time-dependent hydrodynamical simulations in \texttt{MESA} to resolve the non-linear pulsation cycles. 

\section{Numerical Methods} \label{sec:methods}

\begin{figure*}[ht!]
\centering
\includegraphics[width=\linewidth]{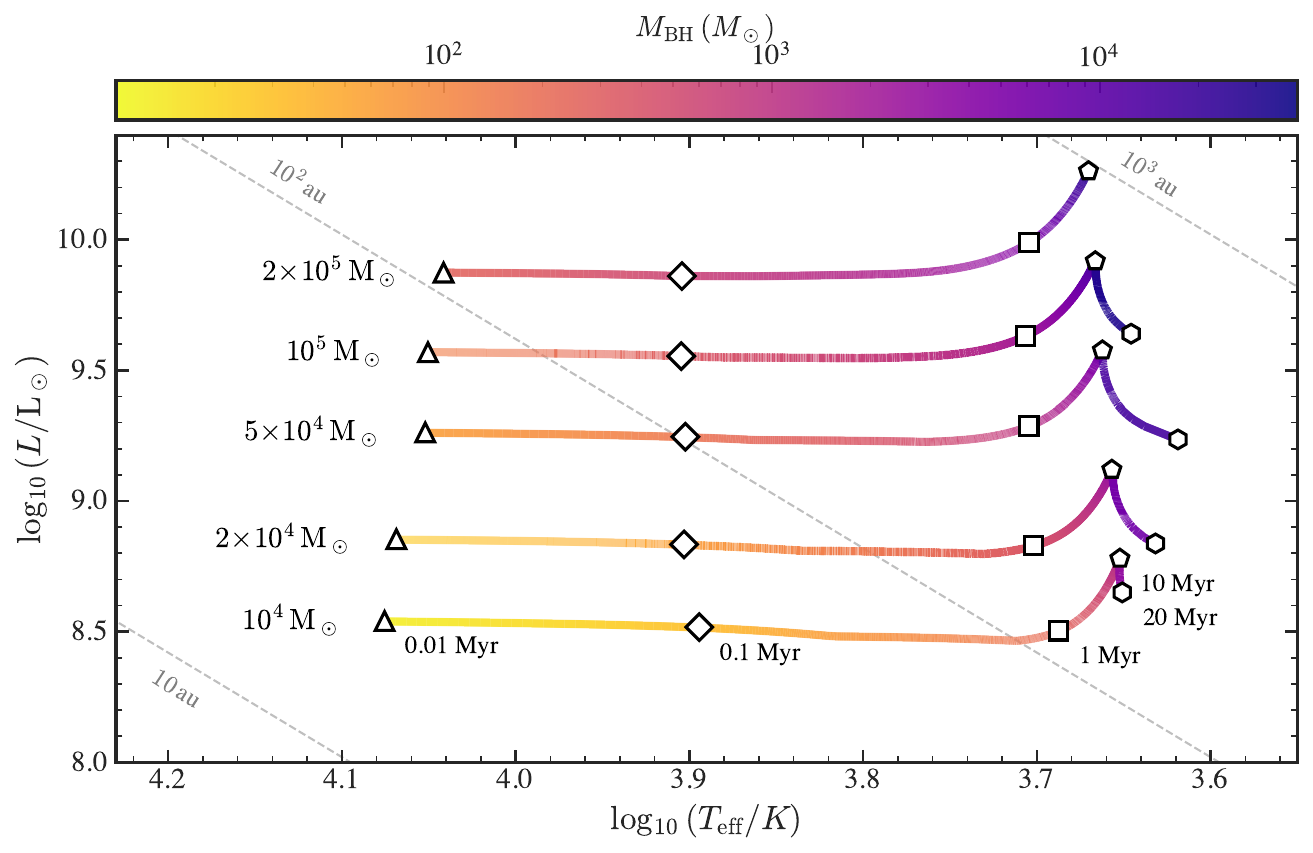}
\caption{Hertzsprung--Russell diagram evolution for our quasi-star models. Tracks show models with different total masses and a central black hole growing from an initial seed of $M_{\rm BH} = 100\,\mathrm{M}_\odot$ to $M_{\rm BH} \approx 0.34\,M_\star$ via accretion. The two highest mass models terminate a bit earlier due to numerical issues. The line color indicates the instantaneous black hole mass, while white markers denote selected evolutionary times. Dashed lines indicate constant stellar radii.}
\label{fig:hrd}
\end{figure*}

\subsection{\texttt{MESA} Quasi-star Models}
We construct a grid of non-rotating quasi-star models with total masses $M_\star$ ranging from $10^4 \Msol$ to $2\times10^5 \Msol$ using the evolutionary code \texttt{MESA} \citep{Paxton2011, Paxton2013, Paxton2015, Paxton2018, Paxton2019, Jermyn2023}, modified to model quasi-stars (Fig.~\ref{fig:hrd}). Our setup follows the ``Coughlin model'' implementation detailed in \citet{Hassan2025}, without accounting for wind mass loss.
The structure is assumed to consist of an interior region transferring energy at the maximum possible rate achievable by convection, and a convective envelope. The envelope is initialized with primordial composition ($X=0.3, Y=0.7, Z=0$)\footnote{The metallicity of LRDs is likely low, but not zero. When testing $M_\star = 10^5 \; \mathrm{M}_\odot$ models with $Z=10^{-3}$ and $Z=10^{-4}$, we found that our results were not meaningfully affected.}
and allowed to reach thermal equilibrium, after which the black hole (initially with 1/1000 the mass of the system) is ``injected" into the center by altering the inner boundary conditions.

The inner boundary mass and radius are calculated following \citet{Hassan2025}, which uses a modified versions of the equations from \citet{Coughlin2024} and incorporates recent corrections described in \citet{Begelman2025}. Within the scope of these equations, the envelope is assumed to have a fixed and uniform adiabatic index of $\gamma=4/3$. The rate $\dot{M}_\mathrm{in}$ at which mass flows from the base of the envelope to the black hole is calculated with $L=\epsilon \dot{M}_\mathrm{in} c^2$, where $\epsilon$ is an efficiency parameter corresponding to the conversion of mass to radiation. Here, $L$ is assumed to be the Eddington limit of the total system mass, $L_{\mathrm{Edd}}(M_{\star})$; this is calculated by estimating the opacity at the base of the ``inefficient convection layer," as defined in \citet{Hassan2025}. The inner boundary luminosity is then calculated using the inner boundary radius $R_i$ via $L=4\pi\eta R_i^2 pc_s$, where $\eta$ is a convective efficiency parameter for the interior. Both $\epsilon$ and $\eta$ are tunable parameters, which we assume to be $\epsilon=\eta=0.1$ (following \citealt{Ball2011} and \citealt{Coughlin2024}); we leave the exploration of other values as a consideration for future work.

For total masses up to $10^5 \; \mathrm{M}_\odot$, our models consistently achieved black hole growth up to $M_{\rm BH} \approx 0.34 M_\star$, terminating only when the mass of the saturated interior region approached its theoretical limit \citep{Hassan2025}. For masses exceeding $10^5 \; \mathrm{M}_\odot$, our simulations terminate prematurely due to numerical issues. However, we successfully evolved the $2\times 10^5 \; \mathrm{M}_\odot$ model to $M_{\rm BH} \approx 0.20 M_\star$,
covering approximately two-thirds of the quasi-star lifetime. We can use these results to extrapolate the properties of systems with $M_\star \gtrsim 10^6 \; \mathrm{M}_\odot$.

\subsection{\texttt{GYRE} Linear Stability Analysis}
To assess the pulsational stability of our quasi-star \texttt{MESA} models, we calculate non-adiabatic oscillation frequencies and growth rates using version 8.1 of the stellar oscillation code \texttt{GYRE} \citep{Townsend2013}. 
To strictly enforce boundary conditions and resolve the steep gradients near the interior and surface, we enable \texttt{add\_double\_points\_to\_pulse\_data}; we explicitly disable \texttt{add\_center\_point\_to\_pulse\_data} to set a zero radial displacement inner boundary condition \texttt{inner\_bound = `ZERO\_R'} at $R_i > 0$. 

We perform a linear non-adiabatic stability analysis to identify unstable radial modes ($\ell=0$). Given the extremely tenuous and super-Eddington nature of quasi-star envelopes ($L/M_\star \sim \Ledd/M_\star$), we anticipate strong non-adiabatic effects where the thermal timescale is comparable to the dynamical timescale. This regime is conducive to the excitation of both classical $\kappa$-mechanism instabilities \citep[driven by ionization zones; see e.g.,][]{Heger1997,YoonCantiello2010} and ``strange modes,'' which are characteristic of objects with high luminosity-to-mass ratios where wave energy is trapped by opacity barriers or density inversions \citep[e.g.,][]{Gautschy1990,Saio1989,Glatzel1994}.

To identify roots in this strongly non-adiabatic regime, we utilize the \texttt{CONTOUR} search method in \texttt{GYRE} \citep{Goldstein2020}. We employ the second-order Magnus Multiple Shooting scheme (\texttt{diff\_scheme = `MAGNUS\_GL2'}) for numerical precision.  
Modes are identified as unstable if the imaginary part of the eigenfrequency is positive ($\Im(\omega) > 0$). By inspecting the differential work ($dW/dx$) and its integral W,
we determine the nature and location of damping and driving of pulsations. 
Finally, we map the boundaries of these unstable solutions to build a theoretical instability strip for LRDs; objects within this strip are expected to exhibit pulsational variability, while those outside should remain stable.

\subsection{Inner Boundary Condition }
A critical component of our analysis is the definition of the inner boundary of the resonant cavity. We adopt the inner radius $R_i$ defined by the \citet{Coughlin2024} model—consistent with our evolutionary calculations in \texttt{MESA}—rather than the formal Bondi radius. This is because linear pulsation codes such as \texttt{GYRE} generally require a background state in hydrostatic equilibrium. In the \citet{Coughlin2024} framework, $R_i$ marks the transition to saturated convection, where the convective flux requires velocities approaching the local sound speed ($v_{\rm conv} \sim c_s$). Consequently, the region $r < R_i$ is dynamically dominated by near-sonic turbulence and the gravitational potential of the black hole, violating the static assumptions required for the background model. Thus, $R_i$ represents the effective ``acoustic floor'' of the hydrostatic envelope.
Moreover, as demonstrated in the analytic derivation by \citet{Hassan2025}, the boundary condition based on the Bondi radius \citep[the ``Ball model'';][]{Ball2012} ceases to exist for evolved quasi-stars. Specifically, the cubic equation governing the inner radius yields no real physical solutions once the black hole mass exceeds a critical threshold ($M_{\rm BH} > M_{\rm crit}$), which occurs at relatively low mass ratios ($M_{\rm BH}/M_{\star} \sim 10^{-2}$).

\section{Results} \label{sec:results}

\subsection{Linear Stability}
We performed a linear non-adiabatic stability analysis on  quasi-star evolutionary tracks using \texttt{GYRE}, focusing on low-order radial ($\ell=0$) modes, where the dominant restoring force is the pressure gradient (p~modes).

\begin{figure}[ht!]
\includegraphics[width=\linewidth]{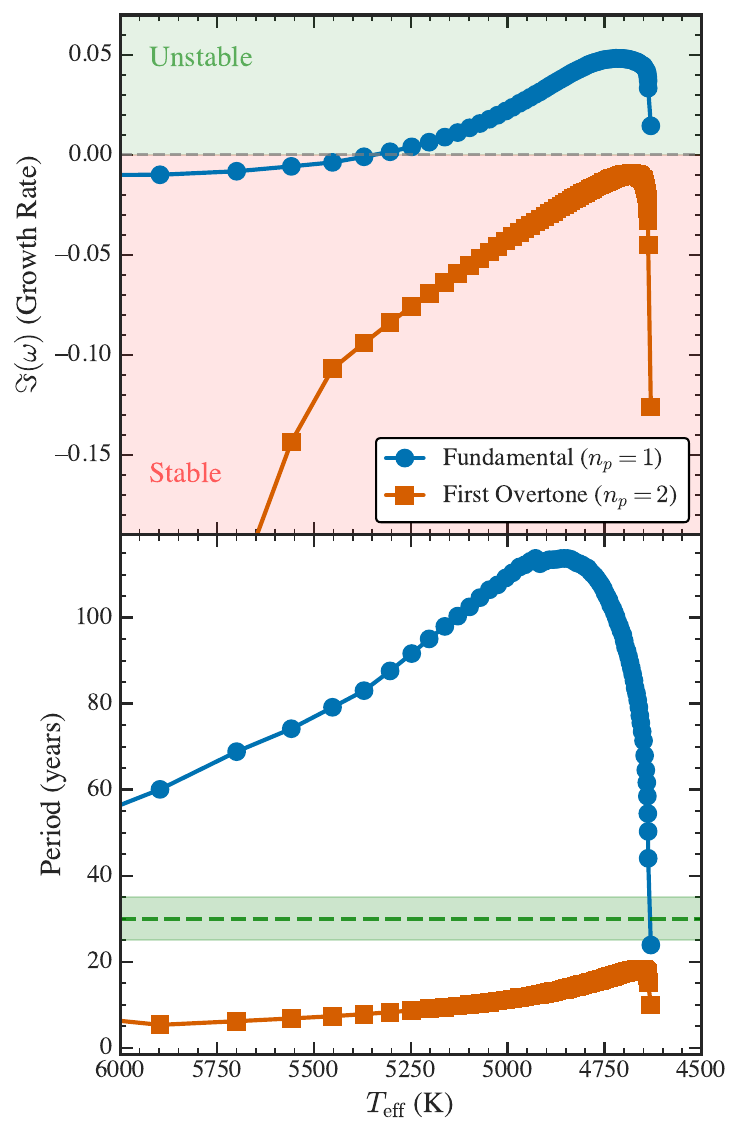}
\caption{\texttt{GYRE} stability analysis for a quasi-star model sequence with $M_\star = 10^5 M_\odot$. {\it Top panel}: Dimensionless growth rate $\Im(\omega)$ as a function of effective temperature for the fundamental mode ($n_{\rm p}=1$, blue circles) and first overtone ($n_{\rm p}=2$, orange squares). The horizontal dashed line marks the stability boundary at $\Im(\omega)=0$. Positive values (green shading) indicate unstable, driven modes, while negative values (red shading) indicate stable, damped modes. The fundamental mode is unstable across most of the temperature range, becoming stable only at $T_{\rm eff} \gtrsim 5300$~K. {\it Bottom panel}: Pulsation period as a function of effective temperature for the same model sequence. The horizontal green band marks the observed $\sim$30-year variability timescale.}
\label{fig:instability}
\end{figure}

Our analysis reveals a clear boundary between stable and unstable regimes in the Hertzsprung--Russell diagram, effectively defining a ``Quasi-Star Instability Strip." For models with effective temperatures $\Teff \gtrsim 5000-5200$\,K, all low-order radial modes are damped ($\Im(\omega) < 0$). However, as  quasi-star models evolve and cool, the damping decreases monotonically, leading to the sequential excitation of modes (Fig.~\ref{fig:instability}).

This theoretical stability boundary is remarkably consistent with the observational dichotomy reported by \citet{Zhang2025}. The source R2211-RX2 (RX2), which shows no significant variability, has an inferred temperature of $\Teff \approx 5000$\,K, placing it on the stable side of our predicted blue edge. In contrast, the variable source RX1, with $\Teff \approx 4000$\,K, lies deep within the unstable region where our models predict vigorous pulsations.

\begin{figure}[ht!]
\includegraphics[width=\linewidth]{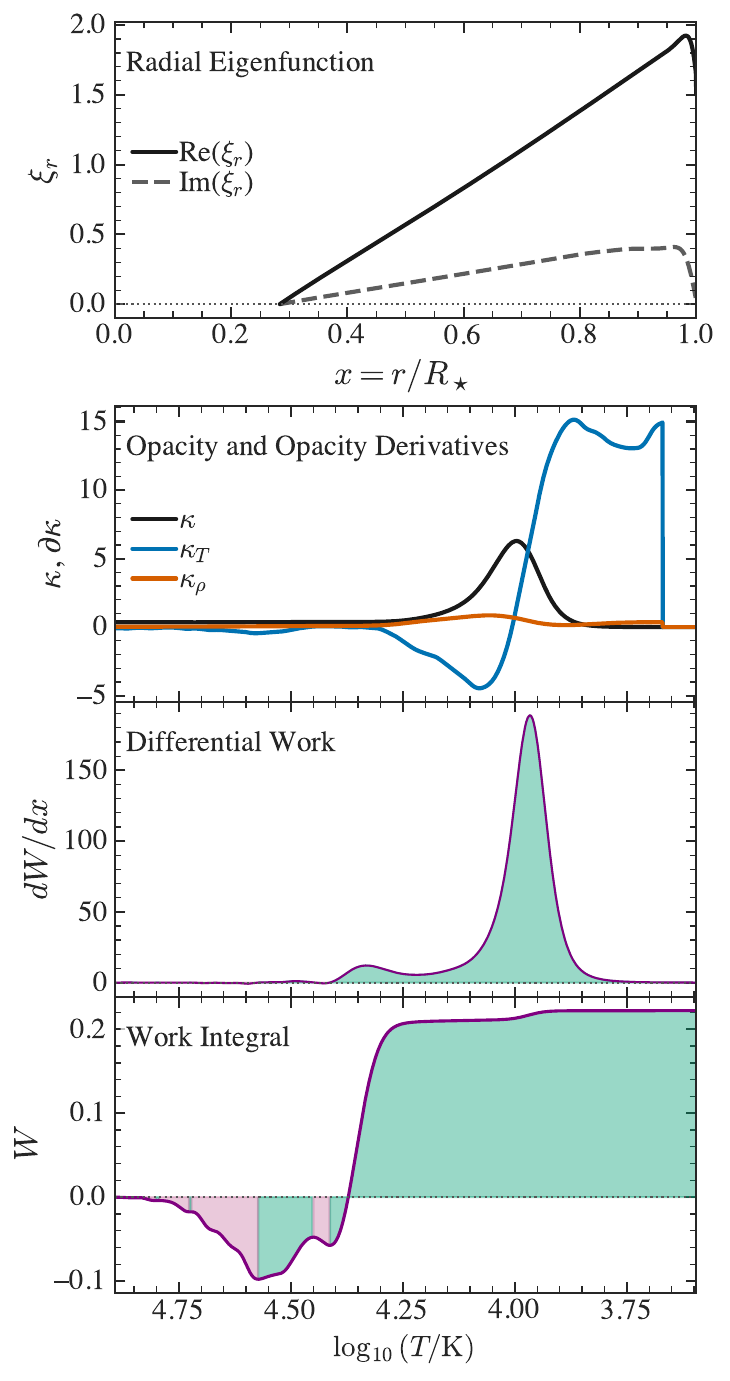}
\caption{Non-adiabatic pulsation analysis of a quasi-star model with mass $2\times10^5 \Msol$ and $\Teff = 4682\,{\rm K}$. {\it Top panel:} radial eigenfunction $\xi_r$ as a function of normalized radius $x = r/R_\star$ for the radial ($\ell =0$) fundamental mode. {\it Bottom panels:} opacity, opacity derivatives $(\partial \ln \kappa/\partial \ln  T)_\rho$ and $(\partial \ln \kappa/\partial \ln\rho)_T$, differential work and its integral W as a function of $\log_{10} T$. Regions of driving  are shown in green, while damping regions are shown in purple. The mode is unstable (W$>$0) and most of the driving occurs at the location of helium recombination around $\log_{10}(T/{\rm K}) \sim 4.4$, with some driving also due to hydrogen recombination at $\log_{10}(T/{\rm K}) \sim 4$. The reported pulsational period for this mode is P = 73.2 yr.}
\label{fig:gyre_stack_fundamental}
\end{figure}

\subsection{Mode Identification}
We identify two distinct p~modes that dominate the variability, appearing in a hierarchical sequence as the envelope expands. 
The fundamental radial mode ($\ell = 0$, $n_{\rm p}=1$) is the first to become unstable crossing $\Teff \approx 5000-5200$\,K. For example, in the model shown in Fig.~\ref{fig:gyre_stack_fundamental}, the fundamental mode is clearly unstable with a positive dimensionless growth rate ($\Im(\omega) \approx +0.039$) and a dimensionless frequency of $\Re(\omega) \approx 0.91$. 

Given the dynamical timescale of the quasi-star, 
\begin{equation}
    \tau_{\rm dyn} =  \sqrt{\frac{R_\star^3}{GM_\star}} \approx 10.6 \,\textrm{yr},
   \label{eq:tdyn}
\end{equation}
we can calculate the pulsational period $P \approx 2\pi\tau_{\rm dyn}/\Re(\omega) = 73.2$ years.
The pulsation growth timescale is $\tau_{\rm growth} \approx \tau_{\rm dyn}/\Im(\omega) = 273$ years. Stability analysis of the evolutionary sequence for a $10^5 \, \Msol$ quasi-star reveals unstable fundamental modes with periods ranging from 30 to 130 years (Fig.~\ref{fig:instability}). This suggests that LRDs inside the instability strip should exhibit secular variability or ``long secondary periods'' spanning decades to centuries. 

\begin{figure}[ht!]
\includegraphics[width=\linewidth]{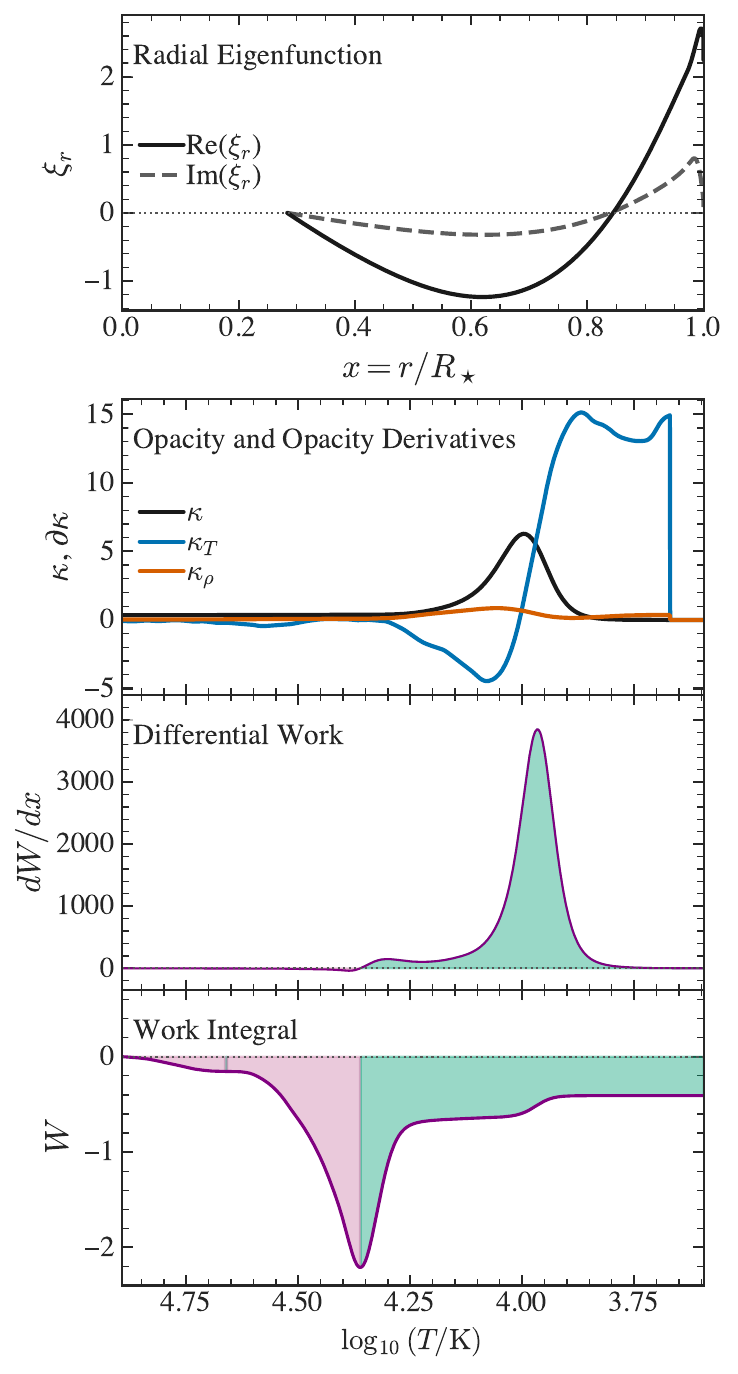}
\caption{Same as Fig.~\ref{fig:gyre_stack_fundamental}, but for the first overtone. This mode is marginally stable according to \texttt{GYRE} non-adiabatic calculations (W$\simeq$0). The predicted pulsation period is P = 21.5 yr. \texttt{MESA} hydrodynamic calculations show this mode to be unstable, with P = 23.5 yr.}
\label{fig:gyre_stack_fundamental_overtone}
\end{figure}

On the other hand, the first overtone ($\ell = 0$, $n_{\rm p}=2$) is the primary candidate for the more rapid variability observed in the lensed system RX1. We present the analysis of the overtone mode for our fiducial model in Fig.~\ref{fig:gyre_stack_fundamental_overtone}. It has a dimensionless frequency $\Re(\omega) \approx 3.10$, corresponding to a period $P \approx 21.5$ years. This is remarkably close to the $\sim 30$-year periodicity reported by \citet{Zhang2025}, even though this specific model was not fine-tuned to match the exact temperature and luminosity of R2211-RX1. 

While this mode is formally stable in this specific snapshot ($\Im(\omega) \approx -0.021$), the damping rate is vanishing rapidly as the star cools. Moreover, this model is found to be unstable with $P \approx 23.5$ years in a non-linear \texttt{MESA} hydrodynamic calculation (see Sec.~\ref{sec:hydro}). Given the close proximity to the stability boundary, we predict overtone modes to be unstable at slightly lower temperatures and higher luminosities, likely driving the decadal-scale pulsations and hysteresis loops observed in the coolest LRDs.
This hierarchy implies that as a luminous quasi-star evolves across the instability strip, it first pulsates in the fundamental mode before exciting the first overtone, potentially leading to multi-periodic variability or mode switching in the coolest objects.

\subsection{Driving Mechanism}

To identify the physical mechanism driving these pulsations, we examined the non-adiabatic eigenfunctions and the differential work, $dW/dx$, which quantifies the energy gained or lost by the mode at each radial shell over one oscillation cycle.

Fig.~\ref{fig:gyre_stack_fundamental} displays the differential work  $dW/dx$ for the unstable fundamental mode, and Fig.~\ref{fig:gyre_stack_fundamental_overtone} shows the same for the marginally stable overtone mode. Visually, the profile is dominated by a sharp, positive peak in the outer envelope layers where the temperature is $T \sim 10^4$\,K, corresponding to the hydrogen ionization zone. As shown in the opacity profile, this feature coincides with the steep opacity cliff associated with H recombination.

 However, inspection of the work integral W reveals that the driving is not sustained solely by this surface layer. While the H-ionization zone produces large local values of $dW/dx$ due to large opacity derivatives, it is located in the tenuous, outermost layers where non-adiabatic thermal damping is strong. Instead, the substantial net positive work required to overcome global damping is provided by the deeper, broader second helium ionization zone ($T \sim 4 \times 10^4$\,K). Although the local opacity derivatives here are more modest, this zone encompasses a significantly larger fraction of the envelope mass and radius. Consequently, it performs the majority of the integrated mechanical work on the envelope.

These results indicate that quasi-stars are driven primarily by the $\kappa$-mechanism in the He~II ionization zone. 
This occurs despite their extremely cool effective temperatures and metal-poor composition, which preclude driving by the deep iron opacity bump ($T \sim 2 \times 10^5$\,K) found in massive stars. Consequently, their pulsational physics aligns more closely with supermassive analogs of classical Cepheids rather than purely H-driven Mira variables \citep{Heger1997,YoonCantiello2010}.

\subsection{The Quasi-Star Instability Strip}\label{sec:instability_strip}
Fig.~\ref{fig:instability_strip} presents the linear stability analysis for our grid of quasi-star models spanning masses from $10^4 \, \Msol$ to $2 \times 10^5 \, \Msol$. We identify a clear, well-defined instability domain for the fundamental radial mode ($\ell=0, n_{\rm p}=1$) on the Hertzsprung-Russell diagram, bounded by a sharp ``Blue Edge'' at high effective temperatures and extending redward toward the Hayashi track.
The hot (blue) edge is located at $\log_{10}(T_{\rm eff}/\mathrm{K}) \approx 3.71$ ($T_{\rm eff} \approx 5000-5200$\,K). Models hotter than this boundary (depicted as white markers in Fig.~\ref{fig:instability_strip}) are stable against fundamental radial pulsations. As the quasi-star envelope expands and cools during its evolution, it crosses this threshold and becomes unstable (colored markers). This transition is consistent across the entire mass range explored, suggesting that the driving mechanism is robust and largely independent of the total stellar mass. The instability domain covers the cool, luminous portion of the HR-diagram where quasi-stars spend the majority of their lifetimes (see Fig.~\ref{fig:hrd}).

Inside the instability strip, the pulsation period for the fundamental mode ranges from $\sim 20$ years for the lowest mass models ($10^4 \, \Msol$) to over $180$ years for the most massive and luminous cases in our grid ($2 \times 10^5 \, \Msol$). We expect these periods to be even longer for quasi-stars with larger masses. The colors in Fig.~\ref{fig:instability_strip} illustrate the period distribution, with solid black contours marking iso-period lines.  The period dependence follows the expected behavior for radial pulsations, shaped by the competing effects of envelope expansion and black hole growth: At a fixed temperature, more luminous (and thus more massive) quasi-stars possess larger radii and lower mean densities, resulting in longer dynamical timescales (Eq.~\ref{eq:tdyn}) and longer pulsation periods. As a quasi-star evolves at roughly constant luminosity toward lower effective temperatures (moving rightward on the diagram), the photospheric radius increases ($R_\star \propto T_{\rm eff}^{-2}$), which primarily drives the period to lengthen. However, this evolutionary excursion is coupled with the growth of the central black hole, which forces the inner envelope boundary $R_{i}$ outward. This inner expansion reduces the effective radial extent of the acoustic cavity, partially counteracting the period lengthening caused by the growth of the photosphere. 

To quantify the impact of this effect, we start from the baseline relation assuming the acoustic cavity extends to the center. As the central black hole grows, the inner boundary of the hydrostatic envelope $R_i$ (the Coughlin radius) moves outward, eventually encompassing a significant fraction of the quasi-star ($R_i/R_\star \gtrsim 0.5$). This drastically reduces the path length for acoustic waves.

For p~modes the pulsation period is proportional to the sound crossing time of the envelope:
\begin{equation}
    P \approx 2 \int_{R_i}^{R_\star} \frac{dr}{c_s(r)}.
    \label{eq:sound_crossing_time}
\end{equation}
To evaluate this analytically, we approximate the structure of the convective envelope as a polytrope. This is a valid approximation given that the envelope is radiation-pressure dominated, corresponding to a polytropic index of $n=3$ ($\gamma \approx 4/3$). In the outer layers, the sound speed profile derived from hydrostatic equilibrium is:
\begin{equation*}
    c_s(r) \approx c_{s,0} \left( 1 - \frac{r}{R_\star} \right)^{1/2},
\end{equation*}
where $c_{s,0} \propto (GM_\star/R_\star)^{1/2}$ represents the characteristic virial sound speed. Substituting this profile into Eq.~\ref{eq:sound_crossing_time} and integrating from the inner boundary ($u_i = 1 - R_i/R_\star$) to the surface ($u=0$), we obtain:
\begin{equation}
    P \propto \frac{R_\star}{c_{s,0}} \left[ 2u^{1/2} \right]_{0}^{1-R_i/R_\star} = \frac{R_\star}{c_{s,0}}  \,\sqrt{1 - \frac{R_i}{R_\star}}.
    \label{eq:P_corrected}
\end{equation}
The geometric factor $(1 - R_i/R_\star)^{1/2}$ becomes significant in the late evolutionary stages of quasi-stars. We note, however, that Eq.~\ref{eq:P_corrected} assumes homologous evolution. In reality, the growth of the inner cavity alters the internal density profile and thermodynamic state of the envelope \citep{Hassan2025}, which likely induces deviations from this simple geometric scaling.

To validate our linear stability predictions, we present  non-linear \texttt{MESA} hydrodynamic calculations in Sec.~\ref{sec:hydro}. These  calculations are represented by star symbols in Fig.~\ref{fig:instability_strip}: The white star, located at $\log_{10} T_{\rm eff} \approx 3.74$, lies outside the instability strip and was confirmed to remain stable in the time-dependent simulations. Conversely, the red star, located deep within the instability domain at $\log_{10} T_{\rm eff} \approx 3.67$, was found to be unstable, developing strong pulsations broadly consistent with the linear predictions. 

While Figure~\ref{fig:instability_strip} focuses on the fundamental mode, our analysis indicates a hierarchical excitation of modes. The first overtone ($\ell=0, n_{\rm p}=2$) is generally stable near the blue edge of the fundamental strip but becomes unstable (or marginally stable) at slightly lower temperatures (see Fig.~\ref{fig:instability}, top panel). These overtone modes have significantly shorter periods ($\sim 10-30$ years, see bottom panel in Fig.~\ref{fig:instability}) and likely drive the decadal-scale variability observed in sources like RX1.

\begin{figure}[ht!]
\includegraphics[width=\linewidth]{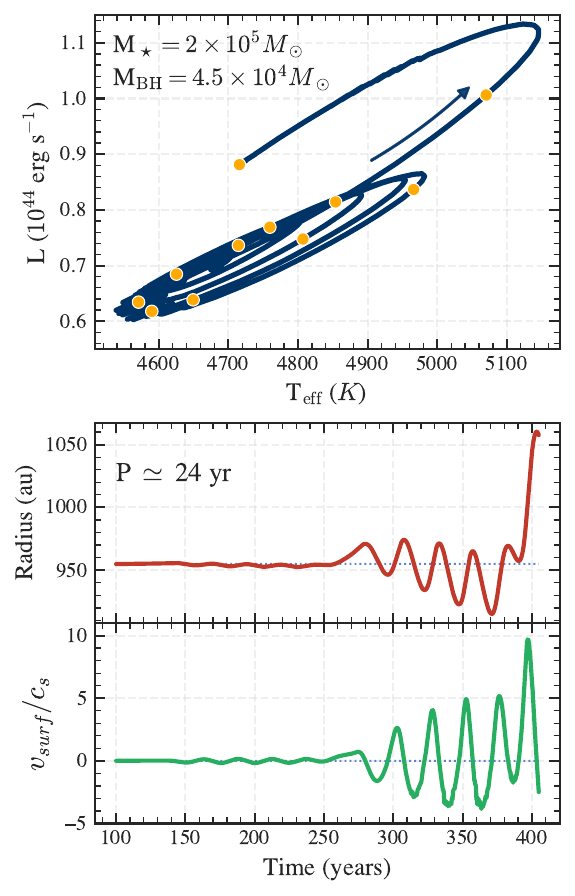}
\caption{Hydrodynamic evolution of a pulsating quasi-star model with envelope mass $M_\star = 2\times10^5\,M_{\odot}$ and black hole mass $M_{\rm BH} = 4.5\times10^4\,M_{\odot}$ (red star symbol in Fig.~\ref{fig:instability_strip}). 
    \textit{Top panel:} The trajectory in the luminosity--temperature plane. The system traces counter-clockwise hysteresis loops, a signature of the thermodynamic ``heat engine'' cycle driving the pulsation. Colored circles along the track mark 10-year intervals during the final 100 years of the simulation.
    \textit{Middle and Bottom panels:} Time evolution of the photospheric radius and the surface Mach number ($v_{\rm surf}/c_s$), respectively. 
    The simulation resolves the rapid growth of an unstable overtone mode ($P \simeq 24$\,yr). The surface velocity eventually becomes highly supersonic, leading to the formation of strong shocks in the envelope. The dotted lines represent results for a model extracted in the stable region (white star symbol in Fig.~\ref{fig:instability_strip}).}
\label{fig:MESA_HYDRO}
\end{figure}

\subsection{\texttt{MESA} Hydrodynamic Calculations}\label{sec:hydro}
We performed hydrodynamic simulations of pulsating quasi-stars by restarting hydrostatic \texttt{MESA} models in evolutionary phases where linear analysis (using \texttt{GYRE}) predicted instability. Upon restarting, we enabled the implicit hydrodynamic solver (\texttt{change\_v\_flag = .true.} and \texttt{new\_v\_flag = .true.}) and restricted the timestep to $\delta t < 0.013$~yr, which is much smaller than the envelope dynamical timescale,  allowing \texttt{MESA} to resolve the onset and growth of pulsations \citep{Paxton2013}.

We focus here on a representative model with high luminosity and low effective temperature ($M_\star = 2\times10^5\,M_{\odot}$, $T_{\rm eff} \approx 4680$\,K). At the analyzed age of $\sim$8\,Myr, the central black hole mass is $M_{\rm BH} \approx 4.5\times10^4\,M_{\odot}$. The top panel in Fig.~\ref{fig:MESA_HYDRO} illustrates the model's evolution on the HR-diagram, alongside the photospheric radius and the surface Mach number ($v_{\rm surf}/c_s$) as a function of time. 

The model exhibits the rapid development of pulsations with a period of $P \approx 24$\,yr, consistent with the period of the marginally stable overtone mode predicted by our linear non-adiabatic \texttt{GYRE} calculations. The simulation eventually terminates when strong shocks develop in the outer quasi-star layers, leading to convergence failures in the implicit solver. 

The trajectory in the luminosity-temperature plane is strikingly similar to the variability reported by \citet{Zhang2025} for the quadruply lensed LRD RX1 (compare their Fig.~3B with the top panel in our Fig.~\ref{fig:MESA_HYDRO}). While RX1 has a luminosity and temperature outside of the range explored by our grid, the very good qualitative agreement between our hydrodynamic tracks and these observations reinforces the interpretation of such LRDs as pulsating quasi-stars. 

\begin{figure*}[ht!]
\centering
\includegraphics[width=\textwidth]{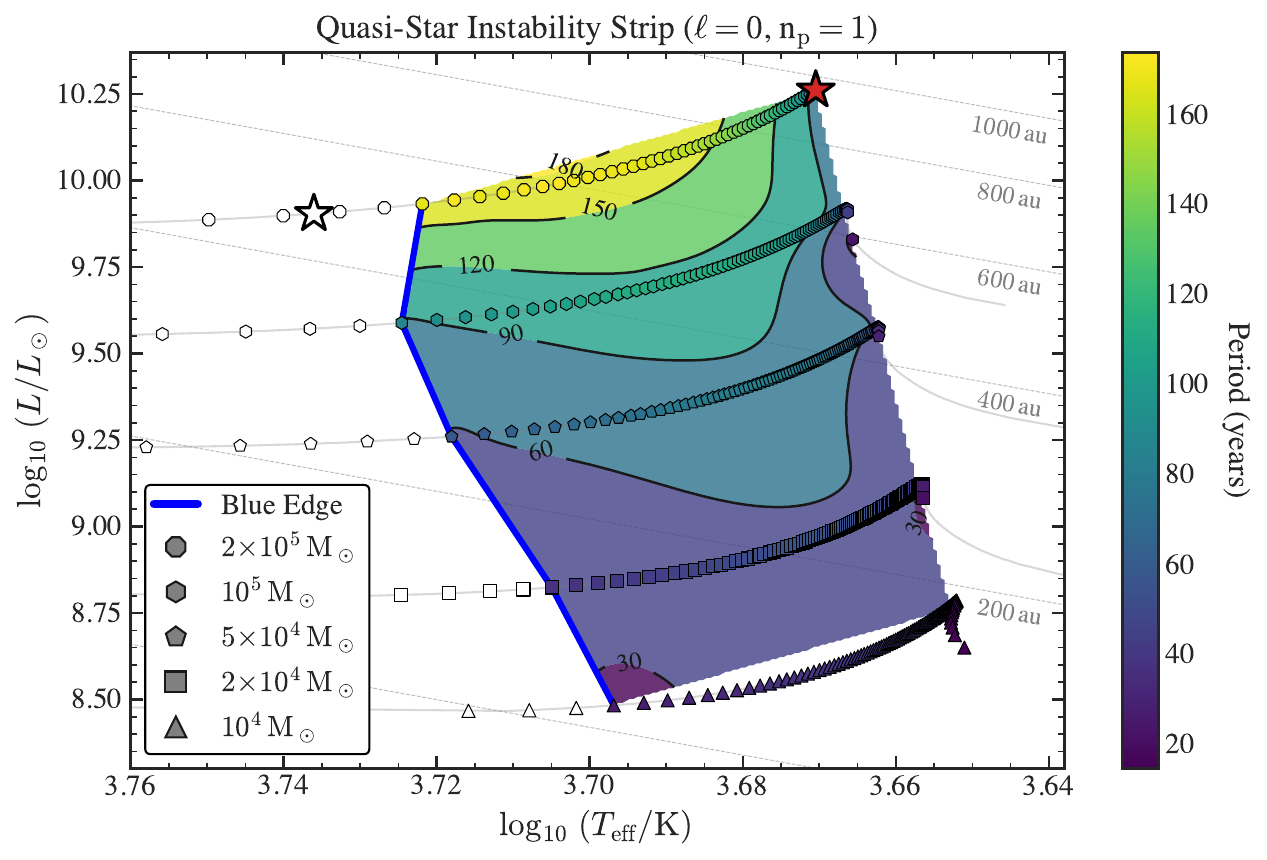}
\caption{ The instability strip for quasi-stars on the Hertzsprung-Russell diagram.  The gray curves represent evolutionary tracks for total stellar masses $M_\star = 10^4, 2\times 10^4, 5\times 10^4, 10^5,$ and $2\times 10^5 \, M_\odot$.  The colored shaded region delineates the instability domain for the fundamental radial mode ($\ell=0, n_{\rm p}=1$), interpolated from the model grid. The color corresponds to the pulsation period in years (see colorbar), with solid black curves marking iso-period contours (labeled in years). The thick blue solid line defines the blue (hot) edge of the instability strip at  $\log_{10} (T_{\rm eff}/\rm{K}) \approx 3.71$. Individual \texttt{GYRE} models are plotted as markers (symbols correspond to quasi-star mass; see legend); colored markers indicate unstable modes, while white markers indicate stable configurations. The star symbols represent the \texttt{MESA} hydro models in the stable (white) and unstable (red) region. Dashed lines indicate constant stellar radii.}
\label{fig:instability_strip}
\end{figure*}

We also tested running models hotter than the blue edge of the instability strip, and found no pulsations in the \texttt{MESA} hydro calculations, consistent with \texttt{GYRE} predictions (dotted flat lines in Fig.~\ref{fig:MESA_HYDRO}, corresponding to white star symbol in Fig.~\ref{fig:instability_strip}).

\section{Discussion} \label{sec:discussion}

The recent discovery of the lensed Little Red Dot RX1 \citep{Zhang2025} provides the first compelling observational counterpart to a pulsating quasi-star phase. Our linear stability analysis confirms that such envelopes are generically unstable to radial pulsations, yielding fundamental modes with periods of $\sim 30$–$180$ years and first-overtone modes with periods of $\sim 10$–$30$ years. The latter is remarkably consistent with the $\sim 30$-year variability cycle (in the co-moving frame) inferred from the $\sim 130$-year baseline provided by the lensed images. Crucially, \citet{Zhang2025} report that RX1 traces a counter-clockwise hysteresis loop in the luminosity–temperature plane, a behavior that is difficult to reconcile with stochastic accretion-driven variability, but is a well-known hallmark of thermodynamic work cycles associated with stellar pulsations. 

We reproduced this behavior by computing hydrodynamic \texttt{MESA} models for quasi-stars both inside and outside the instability region. The unstable models qualitatively match the observed trajectory in the luminosity-temperature plane (Fig.~\ref{fig:MESA_HYDRO}). Conversely, hotter models show no evolution on decadal-to-centennial timescales. This is consistent with the source RX2, which exhibits no significant variability and has an inferred temperature that places it outside the instability strip.

While the consistency between the RX1 and RX2 data and our models is encouraging, significant caution is warranted. The current light curve for RX1 relies on a limited number of epochs (effectively $\sim 4$ independent data points across the century-scale baseline). Although these points trace a coherent loop, the sampling is too sparse to definitively characterize the variability as a stable, periodic pulsation rather than a transient event or quasi-periodic outburst.

Our current analysis relies on several simplifying assumptions. First, our \texttt{GYRE} stability analysis  utilizes hydrostatic snapshots and does not yet capture the full non-linear hydrodynamic coupling between pulsations and mass loss. 
Second, we assume non-rotating, spherical symmetry. In reality, the gas infalling from the host galaxy likely possesses significant angular momentum, and rotation could alter the stability criteria, potentially stabilizing radial modes or exciting non-radial modes not considered here. 

Non-linear \texttt{MESA} hydrodynamic calculations effectively capture the early growth phase of the instability, but they cannot establish the saturation amplitude of the pulsations because \texttt{MESA} does not fully resolve shocks. Shocks develop rapidly during the pulsation cycle and are expected to couple to convection and mass loss; these physical interactions likely determine the final amplitude and period of the variability.

These pulsations and their associated shocks may lead to a rapid increase in the quasi-star mass-loss rate, which in turn could significantly alter the evolutionary trajectory. This phenomenon represents a quasi-star analogue of ``super-AGB'' winds \citep[e.g.][]{YoonCantiello2010}, a scenario that requires self-consistent simulations to determine whether these winds merely strip the envelope gradually or lead to catastrophic disruption. 

A pulsation-driven superwind could drastically shorten the LRD phase by rapidly dispersing the envelope that mediates accretion, thereby limiting the time available for black-hole growth and shaping the final mass of the resulting black-hole seed. However, the net effect of this outflow would depend on its competition with the rapid accretion from the host environment required to form the seed ($\dot{M}_{\rm infall} \gtrsim 0.1 \, \mathrm{M}_{\odot}\,\mathrm{yr}^{-1}$). Since our current models assume constant mass, they do not track this dynamical interplay. In a realistic scenario, the evolutionary path would be determined by the winner of this competition: quasi-stars with moderate infall rates may have their envelopes stripped shortly after crossing the stability ``blue edge'', preventing further growth. Conversely, systems subject to extremely high infall rates may overcome the wind and evolve deeper into the instability strip, where pulsations would likely become even more violent. In either regime, this instability introduces a critical feeedback mechanism that can regulate the maximum mass and lifetime of the LRD phase.  

A critical observational distinction between quasi-stars and classical red supergiants (RSGs) lies in the expected amplitude of short-timescale variability. Hydrodynamic simulations of cool, luminous envelopes \citep[e.g.,][]{Jiang2018} reveal that near the Eddington limit, surface layers are dominated by violent turbulent phenomena, including radiation dominated convection and localized eruptive outbursts. In RSGs, where the pressure scale height is comparable to the stellar radius ($H_P \sim R_{\star}$), these local instabilities subtend a large fraction of the visible disk ($N_{\rm cell} \sim \text{few}$). Consequently, a single localized eruption manifests as a significant photometric fluctuation, producing the large-amplitude stochastic flickering characteristic of RSG light curves.

In contrast, the immense mass of a quasi-star ($M \sim 10^5 \, \mathrm{M}_{\odot}$) confines its atmosphere to a comparatively thin shell ($H_P \sim 1$\,AU $\ll R_{\star} \sim 1000$\,AU). This geometric scaling implies that the photosphere is tiled by $N \sim (R_{\star}/H_P)^2 \sim 10^6$ independent coherent regions. Following standard statistical arguments, the integrated variability from uncorrelated local phenomena (whether convective flickering or localized super-Eddington eruptions) will be suppressed by a factor of $1/\sqrt{N} \sim 10^{-3}$ relative to RSGs. This likely explains why LRD light curves do not exhibit violent short-term variability despite their high luminosities. In this regime, the only mechanisms capable of modulating the total luminosity are coherent, global instabilities ($l=0$), such as the radial pulsations identified in this work.

\section{Conclusions} \label{sec:conclusions}

We have investigated the dynamical stability of quasi-stars (supermassive black hole seeds embedded in massive hydrostatic envelopes) to determine if their pulsational properties can explain the variability observed in the emerging population of Little Red Dots (LRDs). By coupling the stellar evolution code \texttt{MESA} with the non-adiabatic oscillation code \texttt{GYRE}, we have located the blue edge of the theoretical instability strip for these objects and compared it directly to the breakthrough observations of the gravitationally lensed LRDs RX1 and RX2. Our main conclusions are as follows:

\begin{enumerate}
    \item \textbf{The Quasi-Star Instability Strip:} We find that quasi-star envelopes are not inherently unstable; rather, their stability depends critically on their effective temperature. We identify a stability boundary (the ``blue edge") at $\Teff \approx 5000-5200$\,K. Models hotter than this threshold are damped and stable, while cooler models are unstable to strong radial pulsations. 
    This theoretical boundary cleanly separates the two LRDs observed by \citet{Zhang2025}: the stable source RX2 ($\Teff \approx 5000$\,K) is located within (or near) the predicted stable region, while the variable source RX1 ($\Teff \approx 4000$\,K) falls deep within the unstable strip.
    This temperature sensitivity also provides a natural explanation for the stable flux of local LRDs ($T_{\rm eff} \sim 5000$ K) over the ZTF baseline \citep{Burke2025}.

    \item \textbf{Interpretation of Century-Scale Variability:} While numerical limitations prevent the calculation of quasi-stars models in the exact high-luminosity parameter space inferred for RX1, models computed elsewhere in the instability strip exhibit variability cycles in qualitative agreement with the observations. Extrapolating our linear stability analysis to the RX1 regime identifies the First Overtone radial mode ($n_{p}=2$) as the most likely driver of the observed $\sim30$-year variability \citet{Zhang2025}.  The Fundamental mode ($n_{p}=1$) is expected to operate on significantly longer timescales ($\sim$hundred years) and may currently manifest only as secular trends in the light curve. 
    
    \item \textbf{Driving Mechanism:} By analyzing the differential work, we determined that while the hydrogen ionization zone ($T \sim 10^4$\,K) produces a sharp peak in the differential work due to large opacity derivatives, the primary driver of the instability is the $\kappa$-mechanism operating in the deeper helium second ionization zone ($T \sim 4\times 10^4$\,K). Although the local opacity derivatives are smaller here than in the H-ionization zone, the He II zone encompasses a significantly larger mass fraction of the envelope, allowing it to perform greater integrated work over the pulsation cycle. This mechanism physically aligns quasi-stars with supermassive analogs of classical Cepheids.

    \item \textbf{Quasi-Star Mass Loss:} The discovery that quasi-stars are pulsationally unstable could have profound consequences for their evolution. The violent, large-amplitude pulsations predicted by our models will likely drive strong mass loss via a ``super-wind" \citep{YoonCantiello2010}. This mechanism could play an important role in regulating the maximum mass and lifetime of the LRD phase.
\end{enumerate}

These results provide compelling support for the hypothesis that Little Red Dots are indeed quasi-stars, heavy black hole seeds in their final, enshrouded growth phase. The ``smoking gun" for this scenario will be the detection of a statistical sample of variable LRDs. If future monitoring confirms that variability is confined to objects cooler than $\Teff \approx 5000$--$5200$\,K, and periodicities are consistent with the multi-decadal timescales predicted here, it would definitively distinguish the quasi-star model from alternative AGN interpretations.

\begin{acknowledgments}
JBH and RP acknowledge support from NASA award 80NSSC25K7554.
JBH and PJA acknowledge support from award 644616 from the Simons Foundation.  MCB acknowledges support from NASA Astrophysics Theory Program grant 80NSSC24K0940.
The Center for Computational Astrophysics at the Flatiron Institute is supported by the Simons Foundation.
\end{acknowledgments}

\bibliography{biblio}

\end{document}